\documentclass[12pt,prd,tightenlines,nofootinbib,showpacs,showkeys]{revtex4}
\newcommand{\be}{\begin{equation}}
\newcommand{\ee}{\end{equation}}
\usepackage{bm}
\usepackage{graphics}
\usepackage{rotating}
\usepackage{epsfig}
\begin{document}
\title{\begin{flushright}{\rm\normalsize SSU-HEP-11/07\\[5mm]}\end{flushright}
Lamb shift in muonic deuterium atom}
\author{\firstname{A.A.}\surname{Krutov}}
\affiliation{Samara State University, Pavlov street 1, 443011, Samara, Russia}
\author{\firstname{A.P.}\surname{Martynenko}}
\affiliation{Samara State University, Pavlov Street 1, 443011, Samara, Russia}
\affiliation{Samara State Aerospace University named after S.P. Korolyov, Moskovskoye Shosse
34, 443086, Samara, Russia}

\begin{abstract}
We present new investigation of the Lamb shift $(2P_{1/2}-2S_{1/2})$ in
muonic deuterium $(\mu d)$ atom using the three-dimensional quasipotential method
in quantum electrodynamics. The vacuum polarization, nuclear structure and
recoil effects are calculated with the account of contributions of
orders $\alpha^3$, $\alpha^4$, $\alpha^5$ and $\alpha^6$. The results are
compared with earlier performed calculations.
The obtained numerical value of the Lamb shift $202.4139$ meV can be considered as
a reliable estimate for the comparison with forthcoming experimental data.
\end{abstract}

\pacs{36.10.Dr, 31.30.Jv, 12.20.Ds, 32.10.Fn}

\keywords{muonic hydrogen, quantum electrodynamics, Lamb shift}

\maketitle

\section{Introduction}

The muonic deuterium ($\mu d$) is the bound state of negative muon and deuteron.
The lifetime of this simple atom is determined by the muon decay in a time
$\tau_\mu=2.19703(4)\cdot 10^{-6}$ s. When passing from electronic hydrogen
to muonic hydrogen we observe the variation of the relative value of the
nuclear structure and polarizability effects, the electron vacuum
polarization corrections and recoil contributions to the fine
and hyperfine structure of the energy spectrum \cite{BR3,BR4,borie2011,KP1,jentschura3,EGS}.
Muonic atoms represent a unique laboratory for the determination of
the nuclear properties. The experimental investigation of the $(2P-2S)$
Lamb shift in light muonic atoms (muonic hydrogen, muonic deuterium, muonic helium ions)
can give more precise values of the nuclear charge radii \cite{KJ,PSI1,PSI2,CERN,Hauser}.
For more than forty years, a measurement of muonic
hydrogen Lamb shift has been considered one of the fundamental
experiments in atomic spectroscopy. Recently, the progress in muon
beams and laser technology made such an experiment feasible. The
first successful measurement of the ($\mu p$) Lamb shift
transition energy $(2P^{F=2}_{3/2}-2S^{F=1}_{1/2})$ at PSI
(Paul Scherrer Institute) produced the result 49881.88 (76) GHz (206.2949 (32) meV)
\cite{Nature}. It leads to new value of the proton charge radius
$r_p=0.84184(36)(56)$ fm, where the first and second uncertainties
originate respectively from the experimental uncertainty of $0.76$ GHz
and the uncertainty $0.0049$ meV in the Lamb shift value which is
dominated by the proton polarizability term. The new value of proton
radius $r_p$ improves the CODATA value \cite{MT} by an order of
magnitude. Another important project which exists now at PSI
in the CREMA (Charge Radius Experiment with
Muonic Atoms) collaboration proposes to measure several transition
frequencies between $2S$- and $2P$-states in muonic helium ions $(\mu
^4_2He)^+$, $(\mu ^3_2He)^+$ with $50$ ppm precision. As a result new
values of the charge radii of a helion and $\alpha$-particle with
the accuracy 0.0005 fm will be determined. The program of the
investigation of the energy levels in light muonic atoms suggests
that the theoretical calculations of fine and hyperfine structure
of states with $n=1,2$ will be performed with high accuracy. Note
that a discrepancy in the new proton charge radius and CODATA value
induced both a reanalysis of the earlier obtained contributions
to the observed transition frequency and a study of the hypothetical
muon-proton interaction
\cite{jentschura1,jentschura2,carlson,sgk2010}.

Theoretical investigations of the Lamb shift $(2P-2S)$, fine and hyperfine
structure of light muonic atoms was performed many years ago in
Refs.\cite{BR3,BR1,giacomo,JB,BR2,Friar1,Drake} on the basis of the Dirac equation
and nonrelativistic three-dimensional method (see other references in review articles \cite{BR3,EGS}).
Their calculation took into account different QED corrections with the
accuracy 0.01 meV. Recently an approach of \cite{BR3} was extended to
the case of muonic deuterium in \cite{BR4,borie2011} where fine and hyperfine structure
was analyzed with high accuracy. Different corrections to fine and hyperfine
structure of muonic hydrogen are calculated on the basis of three-dimensional method
in quantum electrodynamics in \cite{KP1,KP1999,KP2,M1,M2,M3}.
The vacuum polarization effects of order
$\alpha^5$ were considered in \cite{KN1,KN2,KIKS}. In this work we aim to present new
independent calculation of the Lamb shift $(2P-2S)$ in muonic deuterium $(\mu d)$
with the account of contributions of orders $\alpha^3$, $\alpha^4$,
$\alpha^5$ and $\alpha^6$ on the basis of quasipotential method in
quantum electrodynamics \cite{M1,M2,M3,M4}. We consider such effects of
the electron vacuum polarization, recoil and nuclear structure
corrections which are crucial to attain high accuracy. With the
exception of the nuclear structure and polarizability contribution, we calculate
all corrections in the intervals $(2P_{1/2}-2S_{1/2})$ and $(2P_{3/2}-2P_{1/2})$ with a
precision 0.0001 meV and 0.00001 meV correspondingly. Our purpose consists in a recalculation
and improvement of the earlier obtained results \cite{BR3,BR4} and derivation the
reliable independent estimate for the $(2P_{1/2}-2S_{1/2})$ and $(2P_{3/2}-2S_{1/2})$ Lamb shift,
which can be used for the comparison with forthcoming experimental data. Modern
numerical values of fundamental physical constants are taken from
Ref.\cite{MT}: the electron mass $m_e=0.510998910(13)\cdot 10^{-3}$
GeV, the muon mass $m_\mu=0.1056583668(38)$ GeV, the fine structure
constant $\alpha^{-1}=137.035999084(51)$ \cite{hanneke}, the deuteron mass
$m_d=1.875612793(47)$ GeV. Numerical values
of the proton structure corrections are obtained with the 2010 year CODATA value
for the deuteron charge radius $r_d=2.1424(21)$ fm and $r_d=2.130\pm 0.003\pm 0.009$ fm
from \cite{Sick}.

\section{Effects of vacuum polarization in the one-photon interaction}

Our approach to the investigation of the Lamb shift $(2P-2S)$ in
muonic deuterium is based on the use of quasipotential
method in quantum electrodynamics \cite{M2,M3,M5}, where the two-particle
bound state is described by the Schr\"odinger equation. In perturbation theory the
basic contribution to the muon-deuteron interaction operator is determined by
the Breit Hamiltonian \cite{t4,jentschura3}:
\begin{equation}
H_B=\frac{{\bf p}^2}{2\mu}-\frac{Z\alpha}{r}-\frac{{\bf p}^4}{8m_1^3}-
\frac{{\bf p}^4}{8m_2^3}+\frac{\pi Z\alpha}{2}\left(\frac{1}{m_1^2}+
\frac{\delta_I}{m_2^2}\right)\delta({\bf r})-
\end{equation}
\begin{displaymath}
-\frac{Z\alpha}{2m_1m_2r}\left({\bf p}^2+\frac{{\bf r}({\bf rp}){\bf p}}
{r^2}\right)+\frac{Z\alpha}{r^3}\left(\frac{1}{4m_1^2}+\frac{1}{2m_1m_2}\right)
({\bf L}{\mathstrut\bm\sigma}_1)=H_0+\Delta V^B,
\end{displaymath}
where $H_0={\bf p}^2/2\mu-Z\alpha/r$, $m_1$, $m_2$ are the muon and
deuteron masses, $\mu=m_1m_2/(m_1+m_2)$. The deuteron factor $\delta_I=0$ because
we used further the common definition of the deuteron charge radius $r_d^2=-6\frac{dF_C}{dQ^2}\vert_{Q^2=0}$
\cite{kms,kp1995}.

The wave functions of $2S$- and $2P$-states are equal to
\begin{equation}
\psi_{200}(r)=\frac{W^{3/2}}{2\sqrt{2\pi}}e^{-\frac{Wr}{2}}\left(1-\frac{Wr}{2}
\right),~~~\psi_{21m}(r)=\frac{W^{3/2}}{2\sqrt{6}}e^{-\frac{Wr}{2}}WrY_{1m}(\theta,\phi),
~~W=\mu Z\alpha.
\end{equation}

\begin{figure}
\centering
\includegraphics[width=9.cm]{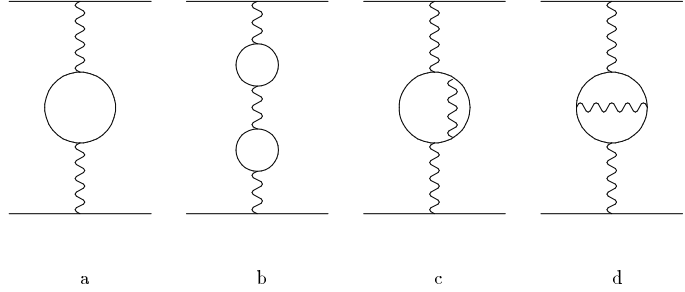}\hspace*{0.4cm}
\caption{Effects of one-loop and two-loop vacuum polarization in
the one-photon interaction.}
\end{figure}

The ratio of the Bohr radius of muonic deuterium to the Compton wavelength of
the electron $m_e/W=0.7$, so, the basic contribution of the electron
vacuum polarization (VP) to the Lamb shift is of order $\alpha(Z\alpha)^2$
(see Fig.1(a)). Accounting for the modification of the Coulomb potential due
to the vacuum polarization in the coordinate representation
\begin{equation}
V^C_{VP}(r)=\frac{\alpha}{3\pi}\int_1^\infty d\xi \rho(\xi)
\left(-\frac{Z\alpha}{r}e^{-2m_e\xi r}\right),~~~
\rho(\xi)=\frac{\sqrt{\xi^2-1}(2\xi^2+1)}{\xi^4},
\end{equation}
we present equations for the one-loop VP contributions to shifts of $2S$-,
$2P$-states and the Lamb shift $(2P-2S)$:
\begin{equation}
\Delta E_{VP}(2S)=-\frac{\mu(Z\alpha)^2\alpha}{6\pi}\int_1^\infty
\rho(\xi)d\xi\int_0^\infty x dx\left(1-\frac{x}{2}\right)^2e^{-x\left(1+
\frac{2m_e\xi}{W}\right)}=-245.3194~meV,
\end{equation}
\begin{equation}
\Delta E_{VP}(2P)=-\frac{\mu(Z\alpha)^2\alpha}{72\pi}\int_1^\infty
\rho(\xi)d\xi\int_0^\infty x^3 dx e^{-x\left(1+
\frac{2m_e\xi}{W}\right)}=-17.6847~meV,
\end{equation}
\begin{equation}
\Delta E_{VP}(2P-2S)=227.6347~meV,
\end{equation}
where we round for definiteness the number to four decimal digits.
The subscript VP designates the contribution of electron vacuum polarization.
Experimental error in a determination of the particle masses and fine
structure constant does not influence on the digits given in (6).
The muon one-loop vacuum polarization correction of order $\alpha(Z\alpha)^4$
is known in analytical form
\cite{EGS}. We included corresponding value $\Delta E_{MVP}(2P-2S)=
\alpha^5\mu^3/30\pi m_1^2=0.01968$ meV to the total shift in section V (Eqs.(71)-(72)).
This result agrees with that in \cite{BR4}.
Two-loop vacuum polarization effects in the one-photon interaction are
shown in Fig.1(b,c,d). To obtain a contribution of the amplitude in Fig.1(b)
to the interaction operator, it is necessary to use the following replacement
in the photon propagator:
\begin{equation}
\frac{1}{k^2}\to\frac{\alpha}{3\pi}\int_1^\infty\rho(\xi)d\xi\frac{1}
{k^2+4m_e^2\xi^2}.
\end{equation}
In the coordinate representation a diagram with two sequential loops
gives the following particle interaction operator:
\begin{equation}
V^C_{VP-VP}(r)=\frac{\alpha^2}{9\pi^2}
\int_1^\infty\rho(\xi)d\xi\int_1^\infty\rho(\eta)d\eta\left(-\frac{Z\alpha}{r}
\right)\frac{1}{(\xi^2-\eta^2)}\left(\xi^2e^{-2m_e\xi r}-\eta^2e^{-2m_e\eta r}
\right),
\end{equation}
where the subscript $(VP-VP)$ corresponds to two sequential loops in the Feynman amplitude
(Fig.1b).
Averaging (8) over the Coulomb wave functions (2), we find the contribution
to the Lamb shift of order $\alpha^2(Z\alpha)^2$:
\begin{equation}
\Delta E_{VP-VP}(2P-2S)=-\frac{\mu\alpha^2(Z\alpha)^2}{18\pi^2}\int_1^\infty d\xi
\int_1^\infty d\eta\frac{\rho(\xi)\rho(\eta)}{(\xi+\eta)}\times
\end{equation}
\begin{displaymath}
\times\left[4m_e^2 W^3\left(4m_e\xi\eta+W(\xi+\eta)\right)\left(8m_e^2 \xi^2\eta^2+4 m_e W\xi\eta (\xi+\eta)+W^2 (\xi^2+\eta^2)\right)\right]=0.2956~meV.
\end{displaymath}

Higher order $\alpha^2(Z\alpha)^4$ correction is determined by an amplitude
with two sequential electron (VP) and muon (MVP) loops. Corresponding
potential is given by
\begin{equation}
\Delta V_{VP-MVP}(r)=-\frac{4(Z\alpha)\alpha^2}{45\pi^2m_1^2}\int_1^\infty
\rho(\xi)d\xi\left[\pi\delta({\bf r})-\frac{m_e^2\xi^2}{r}e^{-2m_e\xi r}\right].
\end{equation}
Its contribution to the shift $(2P-2S)$ is equal to
\begin{equation}
\Delta E_{VP-MVP}(2P-2S)=0.0001~meV.
\end{equation}

\begin{figure}
\centering
\includegraphics[width=9.cm]{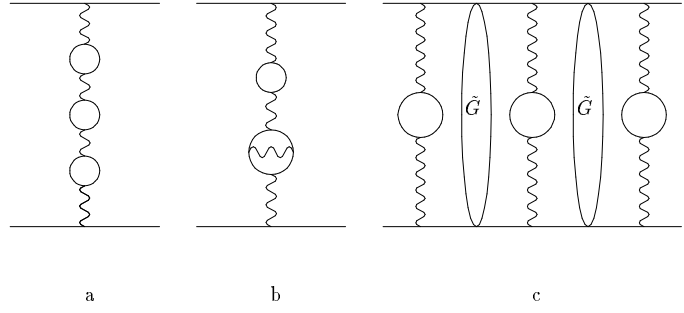}\hspace*{0.4cm}
\caption{Effects of three-loop vacuum polarization in the
one-photon interaction (a,b) and in third order perturbation
theory (c). $\tilde G$ is the reduced Coulomb Green function (33).}
\end{figure}

The two-loop vacuum polarization graphs (Figs.1(b,c,d)) were first calculated in \cite{KS1955,BR1973}.
The particle interaction potential, corresponding to two-loop amplitudes
in Fig.1(c,d) with second order polarization operator, takes the form \cite{KS1955}:
\begin{equation}
\Delta V_{2-loop~VP}^C=-\frac{2}{3}\frac{Z\alpha}{r}
\left(\frac{\alpha}{\pi}\right)^2\int_0^1\frac{f(v)dv}{(1-v^2)}
e^{-\frac{2m_er}{\sqrt{1-v^2}}},
\end{equation}
where the subscript $(2-loop~VP)$ corresponds only to two-loop Feynman amplitudes shown in
Fig.1(c,d), the spectral function
\begin{equation}
f(v)=v\Bigl\{(3-v^2)(1+v^2)\left[Li_2\left(-\frac{1-v}{1+v}\right)+2Li_2
\left(\frac{1-v}{1+v}\right)+\frac{3}{2}\ln\frac{1+v}{1-v}\ln\frac{1+v}{2}-
\ln\frac{1+v}{1-v}\ln v\right]
\end{equation}
\begin{displaymath}
+\left[\frac{11}{16}(3-v^2)(1+v^2)+\frac{v^4}{4}\right]\ln\frac{1+v}{1-v}+
\left[\frac{3}{2}v(3-v^2)\ln\frac{1-v^2}{4}-2v(3-v^2)\ln v\right]+
\frac{3}{8}v(5-3v^2)\Bigr\},
\end{displaymath}
$Li_2(z)$ is the Euler dilogarithm. The potential $\Delta
V^C_{2-loop~VP}(r)$ gives larger contribution as compared with (8)
both to the hyperfine structure and Lamb shift $(2P-2S)$. In the case of the Lamb shift
we find the following contribution:
\begin{equation}
\Delta E_{2-loop~VP}(2P-2S)=1.3704~meV.
\end{equation}
Changing in (12) the electron mass to the muon mass one can obtain two loop
muon vacuum polarization correction. It is known in analytical form from
the paper \cite{Baranger} (we present their result with five decimal digits):
\begin{equation}
\Delta E_{2-loop~ MVP}(2P-2S)=\frac{41}{324}\frac{\alpha^2(Z\alpha)^4\mu^3}{\pi^2 m_1^2}=
0.00017~ meV
\end{equation}
The numerical values of corrections (9), (14) and the desired accuracy of the
calculation show that it is important to consider three-loop
contributions of the electron vacuum polarization (see Fig.2). One part of
corrections to the potential from the diagrams of three-loop vacuum
polarization in the one-photon interaction can be derived by means of
equations (8)-(12) (sequential loops in Fig.2(a,b)) \cite{M3}.
Corresponding contributions to the potential and the splitting
$(2P-2S)$ are given by
\begin{equation}
V^C_{VP-VP-VP}(r)=-\frac{Z\alpha}{r}\frac{\alpha^3}{(3\pi)^3}\int_1^\infty
\rho(\xi)d\xi\int_1^\infty\rho(\eta d\eta\int_1^\infty\rho(\zeta)d\zeta\times
\end{equation}
\begin{displaymath}
\times\left[e^{-2m_e\zeta r}\frac{\zeta^4}{(\xi^2-\zeta^2)(\eta^2-\zeta^2)}
+e^{-2m_e\xi r}\frac{\xi^4}{(\zeta^2-\xi^2)(\eta^2-\xi^2)}+
e^{-2m_e\eta r}\frac{\eta^4}{(\xi^2-\eta^2)(\zeta^2-\eta^2)}\right],
\end{displaymath}
\begin{equation}
V^C_{VP-2-loop~VP}=-\frac{4\mu\alpha^3(Z\alpha)}{9\pi^3}\int_1^\infty
\rho(\xi)d\xi\int_1^\infty\frac{f(\eta)d\eta}{\eta}\frac{1}{r(\eta^2-\xi^2)}
\left(\eta^2e^{-2m_e\eta r}-\xi^2e^{-2m_e\xi r}\right),
\end{equation}
\begin{equation}
\Delta E_{VP-VP-VP}(2P-2S)=0.0005~meV,
\end{equation}
\begin{equation}
\Delta E_{VP-2-loop~VP}(2P-2S)=0.0034~meV,
\end{equation}
where subscripts $(VP-VP-VP)$ and $(VP-2-loop~VP)$ designate only the
Feynman amplitudes shown in Fig.2(a,b) respectively.
Six order vacuum polarization contributions including Eqs.(18)-(19)
were obtained in \cite{BR4}.
The contribution of other diagrams corresponding to the
three loop contribution in $1\gamma$ approximation were calculated in Refs.\cite{KN1,KN2}
for muonic hydrogen.  We estimated their contribution for the Lamb
shift in $(\mu d)$ from the results given in Eqs.(18) and (23) of Ref.\cite{KN1};
the result is  0.0021 meV.  This gives a total three loop contribution
0.0060 meV in one-photon interaction, which is included in Table I.
Two-loop and three-loop vacuum
polarization corrections appearing in second order perturbation theory, are
calculated in the next sections.
Our sum of all three loop VP contributions 0.0086 meV is very close to the total three
loop contribution 0.00842 meV given in Table I of \cite{KIKS} with the account of a rounding.

\begin{figure}
\centering
\includegraphics[width=3.cm]{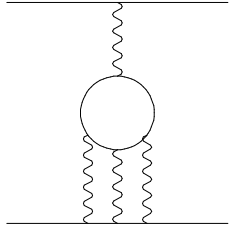}\hspace*{0.2cm}
\caption{The Wichmann-Kroll correction. The wave line shows the Coulomb
photon.}
\end{figure}

Additional one-loop vacuum polarization diagram is presented in
Fig.3. In the energy spectrum it gives the correction of fifth
order in $\alpha$ (the Wichmann-Kroll correction) \cite{WK,MPS}.
The particle interaction potential can be written in this case in
the integral form:
\begin{equation}
\Delta V^{WK}(r)=\frac{\alpha(Z\alpha)^3}{\pi r}\int_0^\infty \frac{d\zeta}
{\zeta^4} e^{-2m_e\zeta r}\left[-\frac{\pi^2}{12}\sqrt{\zeta^2-1}\theta(\zeta-1)
+\int_0^\zeta dx\sqrt{\zeta^2-x^2} f^{WK}(x)\right].
\end{equation}
Exact form of the spectral function $f^{WK}$ is presented in
Refs.\cite{EGS,WK,MPS}. Numerical integration in (20) with the wave
functions (2) gives the following contribution to the Lamb shift:
\begin{equation}
\Delta E^{WK}(2P-2S)=-0.0011~meV.
\end{equation}
This agrees well with other calculations \cite{BR4,KIKS}. The detailed calculation
of all three light-by-light graphs is presented in \cite{KIKS1}. We included in
Table I their estimation using (21) and the results from \cite{KIKS1}.

\section{Relativistic corrections with the vacuum polarization effects}

The electron vacuum polarization effects lead not only to corrections in
the Coulomb potential (3), but also to the modification of other terms
of the Breit Hamiltonian (1). The one-loop vacuum polarization corrections
in the Breit interaction were obtained in \cite{KP1,KP2,jentschura3}:
\begin{equation}
\Delta V^B_{VP}(r)=\frac{\alpha}{3\pi}\int_1^\infty\rho(\xi)d\xi\sum_{i=1}^4
\Delta V_{i,VP}^B(r),
\end{equation}
\begin{equation}
\Delta V_{1,VP}^B=\frac{Z\alpha}{8}\left(\frac{1}{m_1^2}+\frac{\delta_I}{m_2^2}\right)
\left[4\pi\delta({\bf r})-\frac{4m_e^2\xi^2}{r}e^{-2m_e\xi r}\right],
\end{equation}
\begin{equation}
\Delta V_{2,VP}^B=-\frac{Z\alpha m_e^2\xi^2}{m_1m_2r}e^{-2m_e\xi
r}(1- m_e\xi r),
\end{equation}
\begin{equation}
\Delta V_{3,VP}^B=-\frac{Z\alpha}{2m_1m_2}p_i\frac{e^{-2m_e\xi r}}{r}
\left[\delta_{ij}+\frac{r_ir_j}{r^2}(1+2m_e\xi r)\right]p_j,
\end{equation}
\begin{equation}
\Delta V_{4,VP}^B=\frac{Z\alpha}{r^3}\left(\frac{1}{4m_1^2}+\frac{1}{2m_1m_2}
\right)e^{-2m_e\xi r}(1+2m_e\xi r)({\bf L}{\mathstrut\bm\sigma}_1),
\end{equation}
where the superscript B designates the Breit interaction.
In first order perturbation theory (PT) the potentials
$\Delta V_{i,VP}^B(r)$ give necessary contributions of order
$\alpha(Z\alpha)^4$ to the shift $(2P-2S)$:

\begin{figure}
\centering
\includegraphics[width=6.cm]{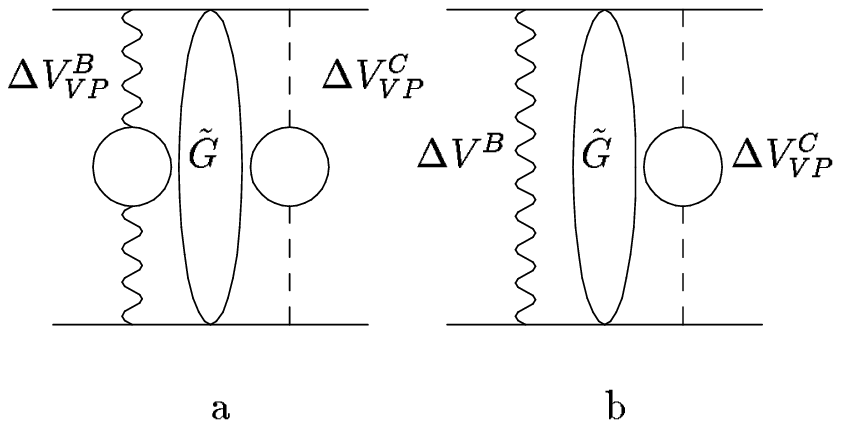}\vspace*{1.0cm}
\includegraphics[width=12.cm]{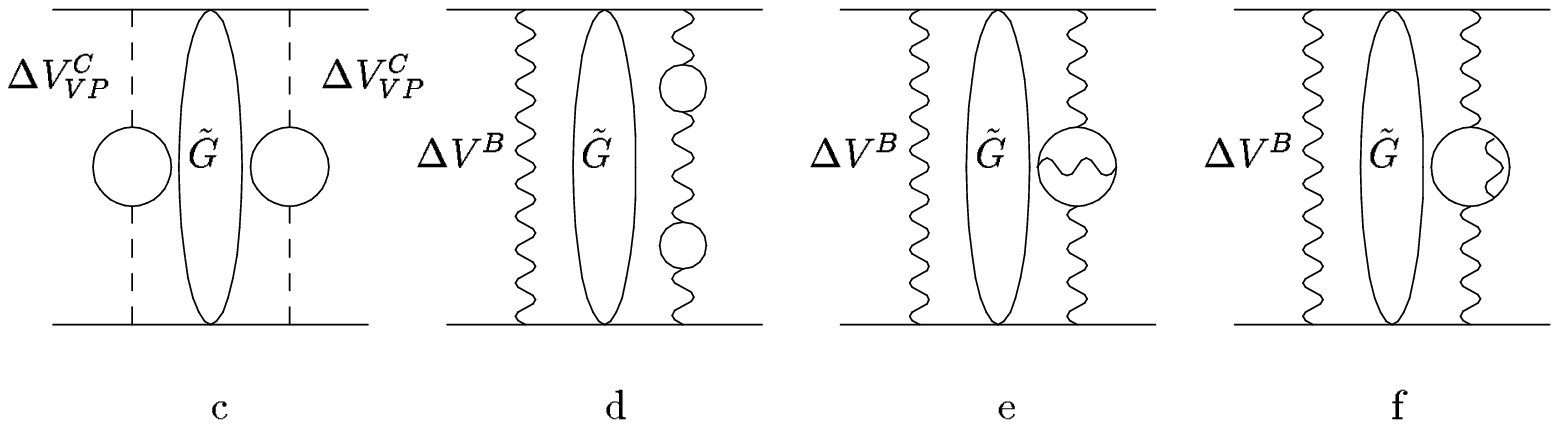}
\caption{Effects of one-loop and two-loop vacuum polarization in
second order perturbation theory (SOPT). The dashed line shows
the Coulomb photon. $\tilde G$ is the reduced Coulomb Green function
(34). The potentials $\Delta V^B$, $\Delta V^C_{VP}$ and $\Delta
V^B_{VP}$ are determined respectively by relations (1), (3) and
(22).}
\end{figure}

\begin{equation}
\Delta E_{1,VP}^B(2P-2S)=-0.0353~meV,
\end{equation}
\begin{equation}
\Delta E_{2,VP}^B(2P-2S)=0.0011~meV,
\end{equation}
\begin{equation}
\Delta E_{3,VP}^B(2P-2S)=0.0012~meV,
\end{equation}
\begin{equation}
\Delta E_{4,VP}^B(2P-2S)=-0.0023~meV.
\end{equation}
The potentials $\Delta V_{2,VP}^B$, $\Delta V_{3,VP}^B$, $\Delta
V_{4,VP}^B$ take into account the recoil effects over the ratio
$m_1/m_2$. We have included in Table I the summary correction of
order $\alpha(Z\alpha)^4$, which is determined by equations
(27)-(30). The next to leading order correction of order
$\alpha^2(Z\alpha)^4$ appears in the energy spectrum from
two-loop modification of the Breit Hamiltonian. We consider
in the potential the term of the leading order in $m_1/m_2$ (the function
$f(v)$ is determined by expression (13)):
\begin{equation}
\Delta V_{2-loop~VP}^B(r)=\frac{\alpha^2(Z\alpha)}{12\pi^2}\left(\frac{1}
{m_1^2}+\frac{\delta_I}{m_2^2}\right)\int_0^1\frac{f(v)dv}{1-v^2}\left[4\pi
\delta({\bf
r})-\frac{4m_e^2}{(1-v^2)r}e^{-\frac{2m_er}{\sqrt{1-v^2}}}\right].
\end{equation}
Corresponding $(2P-2S)$ shift is the following:
\begin{equation}
\Delta E_{2-loop~VP}^B(2P-2S)=-0.0002~meV.
\end{equation}
Other two-loop contributions to the Breit potential are omitted because
they give the energy corrections which lie outside an accuracy of the
calculation in this work.

In second order perturbation theory (SOPT) we have a number of
the electron vacuum polarization contributions in orders
$\alpha^2(Z\alpha)^2$ and $\alpha(Z\alpha)^4$, shown in Fig.4 (b,c):
\begin{equation}
\Delta E_{SOPT}^{VP}=<\psi|\Delta V^C_{VP}\tilde G\Delta V^C_{VP}|\psi>+
2<\psi|\Delta V^B\tilde G\Delta V^C_{VP}|\psi>.
\end{equation}
The abbreviation SOPT is used further in Table I,II for the contributions obtained
in second order PT.
The second order perturbation theory corrections in the energy spectrum of
hydrogen-like system are determined by the reduced Coulomb Green function
$\tilde G$ (RCGF). It has a partial wave expansion \cite{VP}:
\begin{equation}
\tilde G_n({\bf r}, {\bf r'})=\sum_{l,m}\tilde g_{nl}(r,r')Y_{lm}({\bf n})
Y_{lm}^\ast({\bf n'}).
\end{equation}

\begin{figure}
\centering
\includegraphics[width=9.cm]{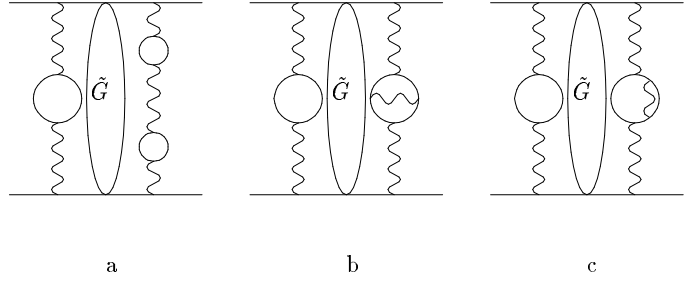}\hspace*{0.4cm}
\caption{The three-loop vacuum polarization corrections in
second order perturbation theory. $\tilde G$ is the reduced Coulomb
Green function.}
\end{figure}

The radial function $\tilde g_{nl}(r,r')$ was presented in \cite{VP} in
a form of the Sturm expansion in the Laguerre polynomials. For a calculation
of the Lamb shift $(2P-2S)$ in muonic deuterium it is convenient to use
the compact representation for the RCGF of $2S$- and $2P$-states,
which was obtained in \cite{KP1,Hameka}:
\begin{equation}
\tilde G(2S)=-\frac{Z\alpha\mu^2}{4x_1x_2}e^{-\frac{x_1+x_2}{2}}\frac{1}
{4\pi}g_{2S}(x_1,x_2),
\end{equation}
\begin{equation}
g_{2S}(x_1,x_2)=8x_<-4x^2_<+8x_>+12x_<x_>-26x^2_<x_>+2x^3_<x_>-4x^2_>-
26x_<x^2_>+23x^2_<x^2_>-
\end{equation}
\begin{displaymath}
-x^3_<x^2_>+2x_<x^3_>-x^2_<x^3_>+4e^x(1-x_<)(x_>-2)x_>+4(x_<-2)x_<(x_>-2)x_>
\times
\end{displaymath}
\begin{displaymath}
\times[-2C+Ei(x_<)-\ln(x_<)-\ln(x_>)],
\end{displaymath}
\begin{equation}
\tilde G(2P)=-\frac{Z\alpha\mu^2}{36x^2_1x^2_2}e^{-\frac{x_1+x_2}{2}}\frac{3}
{4\pi}\frac{({\bf x}_1{\bf x}_2)}{x_1x_2}g_{2P}(x_1,x_2),
\end{equation}
\begin{equation}
g_{2P}(x_1,x_2)=24x^3_<+36x^3_<x_>+36x^3_<x^2_>+24x^3_>+36x_<x^3_>+36x^2_<
x^3_>+49x^3_<x^3_>-3x^4_<x^3_>-
\end{equation}
\begin{displaymath}
-12e^{x_<}(2+x_<+x_<^2)x^3_>-3x^3_<x^4_>+12x_<^3x_>^3[-2C+Ei(x_<)-\ln(x_<)-
\ln(x_>)],
\end{displaymath}
where $x_<=min(x_1,x_2)$, $x_>=max(x_1,x_2)$, $C=0.57721566...$ is
the Euler constant. As a result the two-loop vacuum polarization
contribution to the first term of (33) can be presented
originally in the integral form (Fig.4(c)). The subsequent numerical
integration gives the following results:
\begin{equation}
\Delta E^{VP,VP}_{SOPT}(2S)=-\frac{\mu\alpha^2(Z\alpha)^2}{72\pi^2}
\int_1^\infty\rho(\xi)d\xi\int_1^\infty\rho(\eta)d\eta\times
\end{equation}
\begin{displaymath}
\times\int_0^\infty\left(1-\frac{x}{2}\right)e^{-x\left(1+\frac{2m_e\xi}{W}
\right)}dx\int_0^\infty\left(1-\frac{x'}{2}\right)
e^{-x'\left(1+\frac{2m_e\eta}{W}\right)}dx'g_{2S}(x,x')=-0.1750~meV,
\end{displaymath}
\begin{equation}
\Delta E^{VP,VP}_{SOPT}(2P)=-\frac{\mu\alpha^2(Z\alpha)^2}{7776\pi^2}\int_1^\infty
\rho(\xi)d\xi\int_1^\infty\rho(\eta)d\eta\times
\end{equation}
\begin{displaymath}
\times\int_0^\infty e^{-x\left(1+\frac{2m_e\xi}{W}\right)}dx
\int_0^\infty e^{-x'\left(1+\frac{2m_e\eta}{W}\right)}
dx'g_{2P}(x,x')=-0.0030~meV,
\end{displaymath}
where the superscript $(VP,VP)$ designates the second order PT contribution
when each of the perturbation potentials contains VP correction. The results (39), (40)
agree with the calculation in \cite{KIKS}.
Changing one electron VP potential by the muon VP potential we find that
corresponding correction to the Lamb shift is very small:
\begin{equation}
\Delta E^{VP,MVP}_{SOPT}(2P-2S)=0.0001~meV.
\end{equation}
The second term in (33) has the similar structure (see Fig.4(b)). A
transformation of different matrix elements entering in it is carried out
with the use of algebraic relations of the form:
\begin{equation}
<\psi|\frac{{\bf p}^4}{(2\mu)^2}{\sum}'_m\frac{|\psi_m><\psi_m|}{E_2-E_m}
\Delta V^C_{VP}|\psi>=<\psi|(E_2+\frac{Z\alpha}{r})(\hat H_0+
\frac{Z\alpha}{r}){\sum}'_m\frac{|\psi_m><\psi_m|}{E_2-E_m}\Delta
V_{VP}^C |\psi>=
\end{equation}
\begin{displaymath}
=<\psi|\left(E_2+\frac{Z\alpha}{r}\right)^2\tilde G\Delta V_{VP}^C|\psi>-
<\psi|\frac{Z\alpha}{r}\Delta V_{VP}^C|\psi>+<\psi|\frac{Z\alpha}{r}|\psi>
<\psi|\Delta V_{VP}^C|\psi>.
\end{displaymath}
Omitting further details of the calculation of numerous matrix
elements in (42), we present here the summary numerical
contribution from second term in (33) to the shift $(2P-2S)$:
\begin{equation}
\Delta E^{B,VP}_{SOPT}(2P-2S)=0.0530~meV.
\end{equation}

Other contributions of second order PT (see
Fig.4(d,e,f)) have the general structure similar to Eqs.(39),
(40). They appear after the replacements $\Delta V_{VP}^C\to \Delta
V^B$ and $\Delta V^C_{VP}\to \Delta V^C_{VP,VP}$ in the basic
amplitude shown in Fig.4(c). The estimate of this contribution of
order $\alpha^2(Z\alpha)^4$ to the shift $(2P-2S)$ can be derived if
we take into account in the Breit potential the leading order term
in the ratio $m_1/m_2$. Its numerical value is
\begin{equation}
\Delta E_{SOPT}^{VP,VP;\Delta V^B}(2P-2S)=0.0004~meV.
\end{equation}
The two-loop vacuum polarization contribution is determined also by
the amplitude in Fig.4(a). To obtain its numerical value in the
energy spectrum we have to use Eqs.(3) and (22). In the leading
order in the ratio $m_1/m_2$ we take again the potential (22), which
leads to very small correction of order $\alpha^2(Z\alpha)^4$:
\begin{equation}
\Delta E_{SOPT}^{VP,\Delta V^B_{VP}}(2P-2S)=-0.00001~meV.
\end{equation}

\begin{figure}
\centering
\includegraphics[width=8.cm]{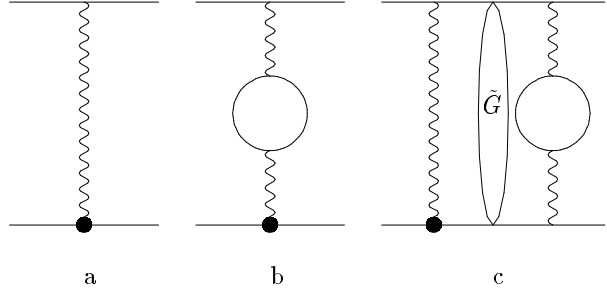}\hspace*{0.4cm}
\caption{The leading order nuclear structure and vacuum polarization
corrections. The thick point represents the nuclear vertex
operator.}
\end{figure}

Three-loop vacuum polarization contributions to the energy spectrum
in second order perturbation theory are presented in Fig.5.
Respective potentials required for their calculation are obtained earlier
in relations (3), (8), (12).
Considering an accuracy of the calculation we can restrict our
analysis by a shift of $2S$-level, which can be written in the form:
\begin{equation}
\Delta E^{VP-VP,VP}_{SOPT}(2S)=-\frac{\mu\alpha^3(Z\alpha)^2}{108\pi^3}
\int_1^\infty\rho(\xi)d\xi\int_1^\infty\rho(\eta)d\eta\int_1^\infty
\rho(\zeta)d\zeta\int_0^\infty dx(1-\frac{x}{2})\times
\end{equation}
\begin{displaymath}
\int_0^\infty dx'(1-\frac{x'}{2})
e^{-x'(1+\frac{2m_e\zeta}{W})}\frac{1}{\xi^2-\eta^2}\left[\xi^2
e^{-x(1+\frac{2m_e\xi}{W})}-\eta^2e^{-x(1+\frac{2m_e\eta}{W})}\right]g_{2S}
(x,x')=-0.0007~meV,
\end{displaymath}
\begin{equation}
\Delta E^{2-loop~VP,VP}_{SOPT}(2S)=-\frac{\mu\alpha^3(Z\alpha)^2}{18\pi^3}
\int_0^1\frac{f(v)dv}{1-v^2}\int_1^\infty\rho(\xi)d\xi\times
\end{equation}
\begin{displaymath}
\times \int_0^\infty
dx\left(1-\frac{x}{2}\right)e^{-x(1+\frac{2m_e}{\sqrt{1-v^2}W})}
\int_0^\infty
dx'\left(1-\frac{x'}{2}\right)e^{-x'(1+\frac{2m_e\xi}{W})}g_{2S}(x,x')
=-0.0018~meV,
\end{displaymath}

In third order perturbation theory (TOPT) the three-loop VP contribution to the Lamb shift
consists of two terms. One part of it is shown in Fig.2(c). This contribution can
be calculated by means of (3), (35)-(38) \cite{KN1,KIKS}. We carry out the coordinate integration
analytically and the integration over three spectral parameters numerically. The result
\begin{equation}
\Delta E_{TOPT}^{VP,VP,VP}(2P-2S)=0.0001~meV
\end{equation}
is in the agreement with \cite{KN1,KIKS}.

\section{Nuclear structure and vacuum polarization effects}

An influence of nuclear structure on the muon motion in muonic deuterium
is determined in the leading order by the root mean square (rms) radius
of the deuteron (charge radius). We present further all charge radius corrections
at two values of $r_d$: $r_d=2.1424(21)$ fm (CODATA 2010) and $r_d=2.130(9)$ fm \cite{Sick}(Fig.6(a)):
\begin{equation}
\Delta E_{str}(2P-2S)=-\frac{\mu^3(Z\alpha)^4}{12}r^2_d=-6.07313\cdot r^2_d=-27.8749(-27.5532)~meV.
\end{equation}
At this point and further the subscript $str$ designates the structure correction.
The precise value of the deuteron charge radius is needed for
the interpretation of new data on transitions in muonic deuterium atom.

\begin{figure}
\centering
\includegraphics[width=8.cm]{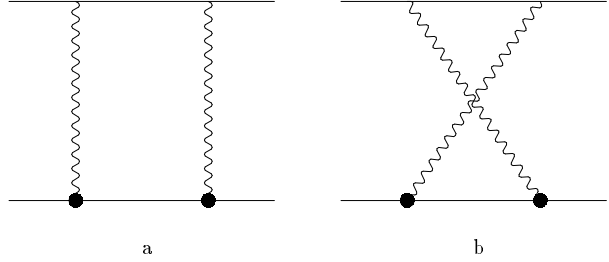}\hspace*{0.4cm}
\caption{Nuclear structure corrections of order $(Z\alpha)^5$.
The thick point is the deuteron vertex operator.}
\end{figure}

There are vacuum polarization corrections connected with the deuteron structure
in first and second orders of perturbation theory (see diagrams in Fig.6(b,c)).
The potential corresponding to the amplitude in Fig.6(b) can be written as follows:
\begin{equation}
\Delta V^{VP}_{str}(r)=\frac{2Z\alpha^2}{9} r^2_d
\int_1^\infty\rho(\xi)d\xi\left[\delta({\bf r})-\frac{m_e^2\xi^2}{\pi r}
e^{-2m_e\xi r}\right].
\end{equation}
Its contribution to the $2P-2S$ Lamb shift is determined by the formula:
\begin{equation}
\Delta E_{str}^{VP}(2P-2S)=-\frac{\mu^3\alpha(Z\alpha)^4}{36\pi}r_d^2
\int_1^\infty\rho(\xi)d\xi\left[1-\frac{16m_e^4\xi^4}{(2m_e\xi+W)^2}\right]=
\end{equation}
\begin{displaymath}
=-0.01350\cdot r_d^2=-0.0620(-0.0612)~meV,
\end{displaymath}

The contribution of the same order $\alpha(Z\alpha)^4$ is specified by the
amplitude in the second order perturbation theory in Fig.6(c):
\begin{equation}
\Delta E^{VP}_{str,SOPT}(2P-2S)=-\frac{\mu^3\alpha(Z\alpha)^4}
{36\pi}r_d^2\int_1^\infty\rho(\xi)d\xi\times
\end{equation}
\begin{displaymath}
\times\frac{-12 + 23 b_1 - 8 b_1^2 - 4 b_1^3 + 4 b_1^4 +
 4 b_1 (3 - 4 b_1 + 2 b_1^2) \ln b_1}{b_1^5}=
\end{displaymath}
\begin{displaymath}
=-0.020487\cdot r_d^2~meV=-0.0940(-0.0929)~meV, ~~~b_1=1+\frac{2m_e}{W}\xi.
\end{displaymath}
Factorizing $r_d^2$ in expressions (49), (51)-(52) we obtain the finite size
correction in the form:
\begin{equation}
\Delta E_{str}(2P-2S)+\Delta E_{str}^{VP}(2P-2S)+\Delta E^{VP}_{str,SOPT}(2P-2S)=-6.10712\cdot r_d^2=
-28.0309(-27.7074)~meV.
\end{equation}

The next important correction of order $(Z\alpha)^5$ is described
by one-loop exchange diagrams (Fig.7). An investigation of elastic
contribution to the Lamb shift and the deuteron polarizability contribution
was performed in \cite{Friar1,Friar2,Friar3,rr1,mf2003}. Recently new detailed
calculation of nuclear structure and polarizability corrections which
improves previous theoretical results is presented in \cite{Pachucki2011}.
We have included in Table I the value of the $(2P-2S)$ shift $1.680(16)$ meV
from \cite{Pachucki2011}.

\begin{figure}
\centering
\includegraphics[width=8.cm]{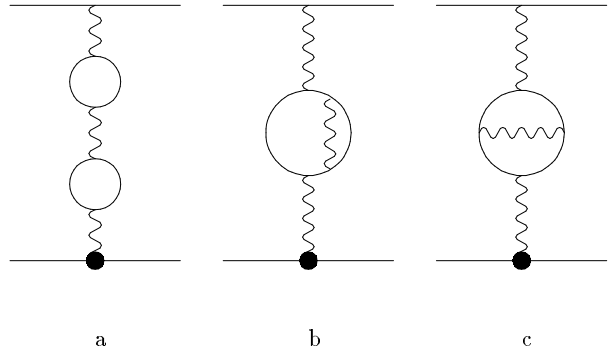}\hspace*{0.4cm}
\caption{Nuclear structure and two-loop vacuum polarization
effects in one-photon interaction. The thick point is the
nuclear vertex operator.}
\end{figure}

Two-loop vacuum polarization corrections with an account of
nuclear structure are presented in Fig.8(a,b,c). The interaction operators
constructed by means of Eq.(7) are determined by integral formulas:
\begin{equation}
\Delta V^{VP-VP}_{str}(r)=\frac{2Z\alpha^3}{27\pi^2} r_d^2
\int_1^\infty\rho(\xi)d\xi\int_1^\infty\rho(\eta)d\eta\times
\end{equation}
\begin{displaymath}
\times\left[\pi\delta({\bf r})-\frac{m_e^2}{r(\xi^2-\eta^2)}\left
(\xi^4 e^{-2m_e\xi r}-\eta^4e^{-2m_e\eta r}\right)\right],
\end{displaymath}
\begin{equation}
\Delta V^{2-loop~VP}_{str}(r)=\frac{4Z\alpha^3}{9\pi^2} r_d^2
\int_0^1\frac{f(v)dv}{1-v^2}\left[\pi\delta({\bf r})-
\frac{m_e^2}{r(1-v^2)}e^{-\frac{2m_er}{\sqrt{1-v^2}}}\right].
\end{equation}
The sum of corrections from (54) and (55) to the Lamb shift $(2P-2S)$ is equal to
\begin{equation}
\Delta E_{str}^{VP,VP}(2P-2S)=-10.5\cdot 10^{-5}\cdot r^2_d=-0.0005(0.0005)~meV.
\end{equation}

\begin{figure}
\centering
\includegraphics[width=10.cm]{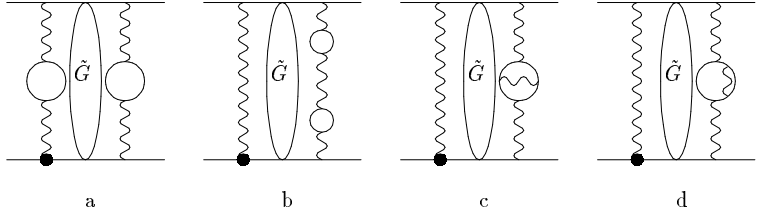}\hspace*{0.4cm}
\caption{Nuclear structure and two-loop vacuum polarization
effects in second order perturbation theory. The thick point is
the nuclear vertex operator. $\tilde G$ is the reduced Coulomb Green
function.}
\end{figure}

Two-loop vacuum polarization and nuclear structure corrections
of order $\alpha^2(Z\alpha)^4$ in second order PT shown in Fig.9(a,b,c,d),
also can be calculated by means of relations discussed in section III. The summary shift
is equal to
\begin{equation}
\Delta E_{str,SOPT}^{VP,VP}(2P-2S)=-9.5\cdot 10^{-5}\cdot r_d^2=-0.0004(0.0004)~meV.
\end{equation}

There exists also the nuclear structure correction of order $\alpha(Z\alpha)^5$
coming from two-photon exchange diagrams with the electron vacuum
polarization insertion (see Fig.10). It can be calculated as the elastic contribution
of order $(Z\alpha)^5$ \cite{mf2003}. However, there is no need to calculate it because
in this case we have the same cancelation between elastic two-photon correction
and deuteron excited states correction as for the contribution of order $(Z\alpha)^5$
\cite{Pachucki2011}. Indeed, using the notations of Ref.\cite{Pachucki2011} we can
present the muon matrix element $P_{VP}$ for nonrelativistic two-photon exchange
with an account of the vacuum polarization in the form:
\begin{equation}
P_{VP}=\frac{2\alpha^3}{3\pi}\phi^2(0)\int_1^\infty\rho(\xi)d\xi\int\frac{d{\bf q}}
{(2\pi)^3}\frac{(4\pi)^2}{q^2(q^2+4m_e^2\xi^2)}\frac{1}{E+\frac{q^2}{2m_1}}
\left[e^{i{\bf q}({\bf R}-{\bf R'})}-1+\frac{q^2}{6}({\bf R}-{\bf R'})^2\right],
\end{equation}
where ${\bf R}$ is the position of the proton with respect to nuclear mass
center. Integrating (58) over $q$ and expanding the resulting expression over
small parameter $\sqrt{2m_1E}|{\bf R}-{\bf R'}|$ we obtain:
\begin{equation}
P_{VP}=\frac{32\alpha^3}{3}m_1\phi^2(0)|{\bf R}-{\bf R'}|^3\int_1^\infty\rho(\xi)d\xi
\biggl[\frac{a_\xi^3-3a_\xi^2+6a_\xi+6e^{-a_\xi}-6}{12a_\xi^4}-
\end{equation}
\begin{displaymath}
-2m_1E|{\bf R}-{\bf R'}|^2
\frac{a_\xi^4-4a_\xi^3+12a_\xi^2-24a_\xi-24e^{-a_\xi}+24}{48a_\xi^6}\biggr],~~~a_\xi=2m_e\xi|{\bf R}-
{\bf R'}|.
\end{displaymath}
It follows from (59) that in the leading order in $\sqrt{2m_1E}|{\bf R}-{\bf R'}|$
elastic correction to atomic energy is canceled by the deuteron excited states correction (see
more detailed discussion in \cite{Pachucki2011}). An estimation of second term contribution
in the square brackets of (59) to the energy spectrum can be derived if we take into
account that the integral over $\xi$ is
determined by the region near $\xi\approx 1$. Expanding second term in (59) at small $a_\xi$ we
obtain $(-\pi/240 a_\xi)$. Then performing analytical integration over $\xi$ and
summing over excited deuteron states we obtain the contribution to the Lamb shift:
\begin{equation}
\delta E^{VP}_{pol}(2P-2S)=-\frac{m_1^2\alpha^3\phi^2(0)}{1024m_e}\biggl[\frac{1}{3}\langle\phi_D|R^2H_DR^2|\phi_D\rangle-
\frac{4}{5}\langle\phi_D|R_iH_DR^2R_i|\phi_D\rangle+
\end{equation}
\begin{displaymath}
+\frac{2}{5}\langle\phi_D|(R_i R_j-\frac{1}{3}\delta_{ij}R^2)H_D
(R_i R_j-\frac{1}{3}\delta_{ij}R^2)|\phi_D\rangle\biggr]=-0.0001~meV,
\end{displaymath}
where $\phi_D$ is the deuteron wave function.
We make all integrations in (60) analytically using the
deuteron wave function in the zero-range approximation \cite{ibk}
\begin{equation}
\phi_D(r)=\sqrt{\frac{\kappa}{2\pi}}\frac{1}{r}e^{-\kappa r},
\end{equation}
where $\kappa=0.0457$ GeV is the inverse deuteron size.

Another term in the Lamb shift of order $\alpha(Z\alpha)^5$ is
determined by muon-line radiative correction to the nuclear size effect.
It was obtained in \cite{Eides_A55} in a suitable form for subsequent numerical estimate:
\begin{equation}
\Delta E_{str}^{\alpha(Z\alpha)^5}(2P-2S)=1.985\frac{\alpha(Z\alpha)^5\mu^3}{8}r_d^2=
9.62\cdot 10^{-4}\cdot r_d^2=0.0044(0.0044)~meV.
\end{equation}
There exists also the correction of order $\alpha(Z\alpha)^5$
with muon vacuum polarization (see diagrams in Fig.(10)).
Accounting for the result of its calculation from \cite{EGS} the total coefficient in (62) should be changed: $1.985\to
1.485$. However, we can consider together with the muon VP and nuclear structure amplitudes in Fig.(10)
the contribution of the deuteron excited states. Calculating this summary contribution
by means of equations similar to (58)-(60) (see also \cite{Pachucki2011})
we observe the cancelation of elastic correction and excited states correction.

\begin{figure}
\centering
\includegraphics[width=7.cm]{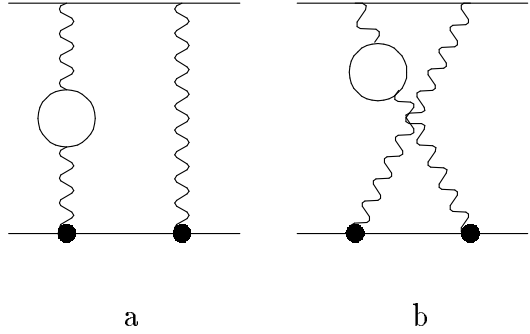}\hspace*{0.4cm}
\caption{The nuclear structure and electron vacuum polarization
effects in the two-photon exchange diagrams. The thick point is the
nuclear vertex operator.}
\end{figure}

Nuclear structure corrections of order $(Z\alpha)^6$ can be derived
with the use of relativistic corrections to nonrelativistic wave
functions in matrix element (49) \cite{EGS,Friar1,friar1997}. We present here
total contribution to the Lamb shift $(2P-2S)$ including additional state independent
correction obtained in \cite{Friar1,friar1997}:
\begin{equation}
\Delta E_{str}^{(Z\alpha)^6}(2P-2S)=\frac{(Z\alpha)^6}{12}\mu^3\Bigl\{r_d^2
\left[\langle\ln \mu Z\alpha r\rangle+C-\frac{3}{2}\right]-\frac{1}{2}r_d^2+
\frac{1}{3}\langle r^3\rangle\langle\frac{1}{r}\rangle-
\end{equation}
\begin{displaymath}
-I_2^{rel}-I_3^{rel}
-\mu^2F_{NR}+\frac{1}{40}\mu^2\langle r^4\rangle\Bigr\}
=-21.28\cdot 10^{-4}\cdot r_d^2+0.0029=-0.0069(-0.0068)~meV,
\end{displaymath}
where the quantities $I_{2,3}^{rel}, F_{NR}$ are written explicitly in \cite{Friar1,friar1997}.
We have extracted in the square brackets the frequently used quantity (main term) for an estimation
of the contribution to the $(2P-2S)$ Lamb shift in hydrogen atom because other corrections are
very small (near 1 $\%$) and could be safely omitted. In the case of muonic deuterium they give
the contribution near $25\%$ of the main term and should be taken into account. 
Separate energy shifts for the $2S$ and $2P$ states are given in \cite{Friar1,friar1997}.
Numerical estimate is obtained on the basis of an exponential parametrization for the charge 
distribution from \cite{Friar1}.

\section{Recoil corrections, muon self-energy and vacuum polarization
effects}

An investigation of different order corrections to the Lamb
shift $(2P-2S)$ of electronic hydrogen has been performed for many
years. Modern analysis of the advances in the solution of this
problem is presented in a review articles \cite{EGS,MPS,SY,SGK}. The most
part of the results was obtained in analytical form, so they can
be used directly in muonic deuterium atom. In this section we
analyze different contributions to the energy spectrum of $(\mu d)$
up to the sixth order in $\alpha$ and derive their numerical
estimations in the Lamb shift $(2P-2S)$.

There are several recoil corrections of different order in $\alpha$
which give important contributions in order to attain the necessary
accuracy of the calculation.
The recoil correction of order $(Z\alpha)^4\mu^3/m_2^2$ to the Lamb shift
appears in the matrix element of the Breit potential with functions (2).
It is calculated for muonic deuterium in \cite{jentschura3,kp1995}:
\begin{equation}
\Delta E_{rec}(2P-2S)=\frac{\mu^3(Z\alpha)^4}{12 m_2^2}=0.0672~meV.
\end{equation}

\begin{figure}
\centering
\includegraphics[width=6.cm]{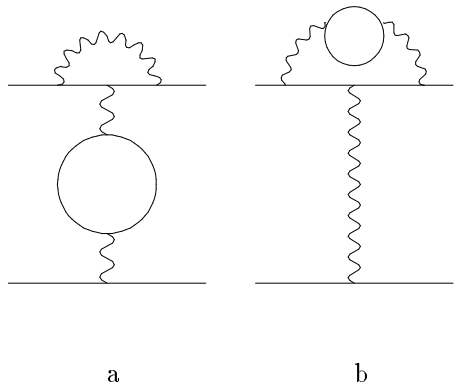}\hspace*{0.4cm}
\caption{Radiative corrections with the vacuum polarization effects.}
\end{figure}

The recoil correction of fifth order in $(Z\alpha)$ is determined
by the expression \cite{SY,EGS}:
\begin{equation}
\Delta E_{rec}^{(Z\alpha)^5}=\frac{\mu^3(Z\alpha)^5}{m_1m_2\pi
n^3}\Bigl[\frac{2}{3}\delta_{l0}
\ln\frac{1}{Z\alpha}-\frac{8}{3}\ln
k_0(n,l)-\frac{1}{9}\delta_{l0}-
\frac{7}{3}a_n-\frac{2}{m_2^2-m_1^2}\delta_{l0}(m_2^2\ln\frac{m_1}{\mu}-
m_1^2\ln\frac{m_2}{\mu})\Bigr],
\end{equation}
where $\ln k_0(n,l)$ is the Bethe logarithm:
\begin{equation}
\ln k_0(2S)=2.811769893120563,
\end{equation}
\begin{equation}
\ln k_0(2P)=-0.030016708630213,
\end{equation}
\begin{equation}
a_n=-2\left[\ln\frac{2}{n}+(1+\frac{1}{2}+...+\frac{1}{n})+1-\frac{1}{2n}\right]\delta_{l0}+
\frac{(1-\delta_{l0})}{l(l+1)(2l+1)}.
\end{equation}
Eq.(65) gives the following numerical correction to the Lamb shift:
\begin{equation}
\Delta E_{rec}^{(Z\alpha)^5}(2P-2S)=-0.0266~meV.
\end{equation}
The recoil correction of the sixth order in $(Z\alpha)$ was
calculated analytically in \cite{EG,Shabaev,Yelkhovsky,M4,PG}:
\begin{equation}
\Delta E_{rec}^{(Z\alpha)^6}(2P-2S)=\frac{(Z\alpha)^6m_1^2}{8m_2}
\left(\frac{23}{6}-4\ln 2\right)=0.0001~meV.
\end{equation}
Omitting explicit form of radiative-recoil corrections of orders
$\alpha(Z\alpha)^5$ and $(Z^2\alpha)(Z\alpha)^4$ from Table 9
\cite{EGS}, we present their numerical value to the Lamb shift
$(2P-2S)$ of muonic deuterium atom:
\begin{equation}
\Delta E_{rad-rec}(2P-2S)=-0.0026~meV.
\end{equation}
The energy contributions obtained in \cite{EG1,EGS,LYE} from radiative corrections to
the lepton line, the Dirac and Pauli form factors and muon
vacuum polarization are given by
\begin{equation}
\Delta E_{MVP,MSE}(2S)=\frac{\alpha(Z\alpha)^4}{8\pi}\frac{\mu^3}{m_1^2}
\Biggl[\frac{4}{3}\ln\frac{m_1}{\mu(Z\alpha)^2}-\frac{4}{3}\ln k_0(2S)+
\frac{38}{45}+
\end{equation}
\begin{displaymath}
+\frac{\alpha}{\pi}\left(-\frac{9}{4}\zeta(3)+\frac{3}{2}
\pi^2\ln 2-\frac{10}{27}\pi^2-\frac{2179}{648}\right)+4\pi Z\alpha\left(
\frac{427}{384}-\frac{\ln 2}{2}\right)\Biggr]=0.7647~meV,
\end{displaymath}
\begin{equation}
\Delta E_{MVP,MSE}(2P)=\frac{\alpha(Z\alpha)^4}{8\pi}\frac{\mu^3}{m_1^2}
\Biggl[-\frac{4}{3}\ln k_0(2P)-\frac{m_1}{6\mu}-
\end{equation}
\begin{displaymath}
-\frac{\alpha}{3\pi}\frac{m_1}{\mu}\left(\frac{3}{4}\zeta(3)-\frac{
\pi^2}{2}\ln 2+\frac{\pi^2}{12}+\frac{197}{144}\right)\Biggr]
=-0.0100~meV.
\end{displaymath}

\begin{table}
\caption{\label{t1}Lamb shift $(2P_{1/2}-2S_{1/2})$ in muonic
deuterium atom.}
\bigskip
\begin{ruledtabular}
\begin{tabular}{|c|c|c|}  \hline
Contribution to the splitting &$\Delta E(2P-2S)$,~meV  & Equation, Reference   \\   \hline
1&2&3 \\  \hline
VP contribution of order $\alpha(Z\alpha)^2$ & 227.6347 & (6), \cite{BR4}  \\
in one-photon interaction &  &   \\   \hline
Two-loop VP contribution of order  $\alpha^2(Z\alpha)^2$&1.6660   &(9), (14), \cite{BR4}    \\
in one-photon interaction &   &   \\    \hline
VP and MVP contribution in  &0.0001     & (11)  , \cite{BR4}      \\
one-photon interaction &     &     \\    \hline
Three-loop VP contribution in  &0.0060    &(17), (18), \cite{KN1,KIKS}     \\
one-photon interaction &    &     \\    \hline
The Wichmann-Kroll correction  & -0.0011  &  (21), \cite{BR4,KIKS}  \\  \hline
Light-by-light contribution &   0.0001    &   \cite{KIKS}   \\   \hline
Relativistic and VP corrections of order  & -0.0353    &  (27)-(30), \cite{jentschura3}     \\
$\alpha(Z\alpha)^4$ in first order PT      &       &    \\    \hline
Relativistic and two-loop VP  & -0.0002    &  (32)    \\
corrections of order $\alpha^2(Z\alpha)^4$   &       &    \\
in first order PT   &       &    \\    \hline
Two-loop VP contribution of order   &0.1720   &(39)-(41) , \cite{KIKS}    \\
$\alpha^2(Z\alpha)^2$ in second order PT &    &    \\    \hline
Relativistic and one-loop VP  & 0.0530    &  (43), \cite{jentschura3}    \\
corrections of order $\alpha(Z\alpha)^4$  &       &    \\
in second order PT  &    &    \\    \hline
Relativistic and two-loop VP   & 0.0004    &  (44)-(45)    \\
corrections of order $\alpha^2(Z\alpha)^4$  &       &    \\
in second order PT  &    &    \\    \hline
Three-loop VP contribution in           &0.0025  &(46)-(47), \cite{KIKS}  \\
second order PT of order $\alpha^3(Z\alpha)^2$  &      &    \\    \hline
Three-loop VP contribution in           &0.0001  &(48), \cite{KN1,KIKS}, \\
third order PT of order $\alpha^3(Z\alpha)^2$  &      &    \\    \hline
Nuclear structure contribution of order $(Z\alpha)^4$   &-27.8749  & (49), \cite{BR4,EGS} \\   \hline
Nuclear structure and polarizability &  &   \\
contribution of order $(Z\alpha)^5$ &1.6800   &\cite{Pachucki2011}  \\   \hline
Nuclear structure and VP contribution  &-0.0620  & (51)   \\
in $1\gamma$ interaction of order $\alpha(Z\alpha)^4$ &   &  \\   \hline
Nuclear structure and VP contribution   &-0.0940  & (52)   \\
in second order PT of order $\alpha(Z\alpha)^4$ &   &  \\   \hline
Nuclear structure and two-loop VP  &-0.0005  & (56)   \\
contribution in $1\gamma$ interaction of order $\alpha^2(Z\alpha)^4$ &   &  \\   \hline
Nuclear structure and two-loop VP contribution&-0.0004  & (57)  \\
in second order PT of order $\alpha^2(Z\alpha)^4$ &   &  \\   \hline
\end{tabular}
\end{ruledtabular}
\end{table}
\begin{table}
Table I (continued).\\
\bigskip
\begin{ruledtabular}
\begin{tabular}{|c|c|c|}  \hline
1&2&3 \\  \hline
Nuclear structure and polarizability & -0.0001 & (60)   \\
contribution of order $\alpha(Z\alpha)^5$     &      &     \\
with VP correction  &    &     \\   \hline
Nuclear structure contribution of order   &0.0044  & (62), \cite{EG1}   \\
$\alpha(Z\alpha)^5$ with muon-line radiative correction &          &    \\   \hline
Nuclear structure contribution   &-0.0069  & (63), \cite{Friar1,friar1997}   \\
of order $(Z\alpha)^6$ &          &    \\   \hline
Recoil correction of order $(Z\alpha)^4$   &0.0672   &(64),~\cite{jentschura3}   \\   \hline
Recoil correction of order $(Z\alpha)^5$   &-0.0266   &(69),~\cite{BR4,SY,EGS}   \\   \hline
Recoil correction of order $(Z\alpha)^6$   &0.0001   &(70),~\cite{EGS}   \\   \hline
Recoil correction to VP of order &  0.0002  &     \cite{jentschura3}  \\
$\alpha(Z\alpha)^5$ (seagull term)  &    &    \\   \hline
Radiative-recoil corrections  &-0.0026 &(71), Table 9 \cite{EGS}   \\
of orders $\alpha(Z\alpha)^5$, $(Z^2\alpha)(Z\alpha)^4$  &   &   \\  \hline
Muon self-energy and MVP contribution &-0.7747 & (72)-(73),~\cite{BR4,EGS}  \\  \hline
Muon form factor $F_1'(0)$, $F_2(0)$ contributions & -0.0018& (78),~\cite{EGS,KP1,BCR}   \\
of order $\alpha^2(Z\alpha)^4$  &    &    \\    \hline
VP correction to muon self-energy &-0.0047 & (80),~\cite{KP1,EGS}  \\  \hline
HVP contribution & 0.0129 & \cite{Friar1999,M6} \\  \hline
Total contribution  & \multicolumn{2}{c|}{$202.4139\pm 0.0573$ ($r_d=2.1424(21)$ fm)}  \\
                  & \multicolumn{2}{c|}{$202.7375\pm  0.2352$ ($r_d=2.130(9)$ fm)} \\  \hline
\end{tabular}
\end{ruledtabular}
\end{table}

The diagram in Fig.11(b) with electron loop polarization insertion in
the radiative photon gives the contribution to the energy spectrum,
which can be expressed in terms of the slope of the Dirac form factor
$F_1'$ and the Pauli form factor $F_2$ \cite{EGS}:
\begin{equation}
\Delta E_{rad+VP}(nS)=\frac{\mu^3}{m_1^2}\frac{(Z\alpha)^4}{n^3}\left[4m_1^2
F_1'(0)\delta_{l0}+F_2(0)\frac{C_{jl}}{2l+1}\right],
\end{equation}
\begin{equation}
C_{jl}=\delta_{l0}+(1-\delta_{l0})\frac{[j(j+1)-l(l+1)-\frac{3}{4}]}{l(l+1)}\frac{m_1}{\mu}.
\end{equation}
Two-loop contribution to the form factors $F_1'(0)$ and $F_2(0)$ was
calculated in \cite{BCR} (see also \cite{BR3,EGS}):
\begin{equation}
m_1^2F_1'(0)=\left(\frac{\alpha}{\pi}\right)^2\left[\frac{1}{9}\ln^2\frac
{m_1}{m_e}-\frac{29}{108}\ln\frac{m_1}{m_e}+\frac{1}{9}\zeta(2)+\frac{395}
{1296}\right],
\end{equation}
\begin{equation}
F_2(0)=\left(\frac{\alpha}{\pi}\right)^2\left[\frac{1}{3}\ln\frac{m_1}{m_e}
-\frac{25}{36}+\frac{\pi^2}{4}\frac{m_e}{m_1}-4\frac{m_e^2}{m_1^2}\ln\frac{m_1}
{m_e}+3\frac{m_e^2}{m_1^2}\right].
\end{equation}
Then the correction to the Lamb shift is equal to
\begin{equation}
\Delta E_{rad+VP}(2P-2S)=-0.0018~meV.
\end{equation}

To estimate the muon self-energy and electron vacuum polarization
contribution in Fig.11(a), we use the relation obtained in \cite{KP1}:
\begin{equation}
\Delta E^{VP}_{MSE}=\frac{\alpha}{3\pi
m_1^2}\ln\frac{m_1}{\mu(Z\alpha)^2} \left[<\psi_n|\Delta\cdot
\Delta V^C_{VP}|\psi_n>+2<\psi_n|\Delta V^C_{VP}\tilde G
\Delta\left(-\frac{Z\alpha}{r}\right)|\psi_n>\right].
\end{equation}
The sum of all matrix elements which appear in Eq.(79) leads to the
following shift $(2P-2S)$:
\begin{equation}
\Delta E^{VP}_{MSE}(2P-2S)=-0.0047~meV.
\end{equation}
The hadron vacuum polarization (HVP) contribution which can be taken into account
on the basis of the numerical result obtained for muonic hydrogen
in \cite{Friar1999,M6} is included in Table I. The error of the measurement of
the cross section $\sigma(e^+e^-\to\pi^+\pi^-)$ was decreased to a few per cents.
So, we estimate in $5\%$ ($\pm 0.0006~meV$) corresponding theoretical error of
HVP correction.

\section{Fine structure of the 2P-state}

The leading order $(Z\alpha)^4$ contribution to fine structure
is determined by the operator $\Delta V^{fs}$:
\begin{equation}
\Delta
V^{fs}(r)=\frac{Z\alpha}{4m_1^2r^3}\left[1+\frac{2m_1}{m_2}+2a_\mu
\left(1+\frac{m_1}{m_2}\right)\right]({\bf
L}{\mathstrut\bm\sigma}_1).
\end{equation}
$\Delta V^{fs}$ includes the recoil correction and muon
anomalous magnetic moment $a_\mu$ correction. Fine structure
interval $(2P_{3/2}-2P_{1/2})$ for muonic deuterium can be
written in the form \cite{lautrup,em2010,ekm2011}:
\begin{equation}
\Delta E^{fs}=E(2P_{3/2})-E(2P_{1/2})=
\end{equation}
\begin{displaymath}
=\frac{\mu^3(Z\alpha)^4}{32 m_1^2}
\left[1+\frac{2m_1}{m_2}+2a_\mu\left(1+\frac{m_1}{m_2}\right)\right]
+\frac{5m_1(Z\alpha)^6}{256}-\frac{m_1^2(Z\alpha)^6}{64m_2}+
\end{displaymath}
\begin{displaymath}
+\frac{\alpha(Z\alpha)^6 \mu^3}{32\pi m_1^2}
\left[\ln\frac{\mu(Z\alpha)^2}{m_1}+\frac{1}{5}\right]+\alpha(Z\alpha)^4A_{VP}
+\alpha^2(Z\alpha)^4B_{VP}+A_{str}(Z\alpha)^6\mu^2\cdot r^2_d.
\end{displaymath}

This expression includes the relativistic correction of order
$(Z\alpha)^6$, which can be calculated on the basis of the Dirac
equation, relativistic recoil effects of order $m_1(Z\alpha)^6/m_2$,
correction of order $\alpha(Z\alpha)^6$
enhanced by the factor $\ln(Z\alpha)$ \cite{EGS}, a number
of terms of fifth and sixth order in $\alpha$ which are determined
by effects of the vacuum polarization and nuclear structure.
Recoil correction $(-m_1^3(Z\alpha)^4/32m_2^2)$ (the Barker-Glover correction
\cite{bg}) is also taken into account in Eq.(82). This is evident from the
expansion of first term in (82) over the mass ratio $m_1/m_2$ up to second order terms:
$m_1(Z\alpha)^4(1-m_1/m_2)/32$.
The contributions to the coefficients $A_{VP}$ and $B_{VP}$ arise in
first and second orders of perturbation theory. Numerical values
of terms in the expression (82), which are presented in
analytical form, are quoted in Table II for a definiteness with an
accuracy $0.00001$ meV. Fine structure interval (82) in the
energy spectrum of electronic hydrogen is considered for a long time
as a basic test of quantum electrodynamics \cite{EGS,SY}.

Fine structure potential with the leading order vacuum polarization
and its contribution to the coefficient $A_{VP}$ are given by (\cite{KP1}):
\begin{equation}
\Delta V^{fs}_{VP}(r)=\frac{\alpha(Z\alpha)}{12\pi
m_1^2r^3}\int_1^\infty\rho(s)ds
[1+\frac{2m_1}{m_2}+2a_\mu(1+\frac{m_1}
{m_2})]e^{-2m_esr}(1+2m_esr)({\bf L}{\mathstrut\bm\sigma}_1),
\end{equation}
\begin{equation}
\Delta E_1^{fs}=\frac{\mu^3\alpha(Z\alpha)^4}{96\pi m_1^2}\left[1+\frac{2m_1}{m_2}+
2a_\mu\left(1+\frac{m_1}{m_2}\right)\right]\int_1^\infty\rho(\xi)d\xi
\frac{1+6\frac{m_e}{W}\xi}{(1+2\frac{m_e}{W}\xi)^3}=
0.00346~~meV.
\end{equation}

Higher order corrections $\alpha^2(Z\alpha)^4$ contributing to
$a_\mu$ are taken into account in this expression as well as
recoil effects. The same order $O(\alpha(Z\alpha)^4)$ contribution can
be obtained in second order perturbation theory in the form:
\begin{equation}
\Delta E^{fs}_{VP,SOPT}=\frac{\alpha(Z\alpha)^4\mu^3}{1728\pi m_1^2}\left[1+2a_\mu+
(1+a_\mu)\frac{2m_1}{m_2}\right]\int_1^\infty\frac{\rho(\xi)d\xi}{(1+2\frac{m_e}{W}\xi)^5}\times
\end{equation}
\begin{displaymath}
\times\left[18\frac{2m_e\xi}{W}\left(\frac{8m_e\xi}{W}+11\right)+4\left(1+\frac{2m_e\xi}{W}\right)
\ln\left(1+\frac{2m_e\xi}{W}\right)+3\right]=0.00229~~meV.
\end{displaymath}
Let us consider two-loop vacuum polarization contributions in the one-photon
interaction shown in Fig.1. They give corrections to fine structure splitting of $P$-wave levels of order
$\alpha^2(Z\alpha)^4$. In the coordinate representation, the interaction operator has the form
\cite{M3,M5}:
\begin{equation}
\Delta V^{fs}_{VP-VP}(r)=\frac{Z\alpha}{r^3}\left[\frac{1+2a_\mu}{4m_1^2}+
\frac{1+a_\mu}{2m_1m_2}\right]({\bf L}{\mathstrut\bm\sigma}_1)\times
\end{equation}
\begin{displaymath}
\times\left(\frac{\alpha}{3\pi}\right)^2\int_1^\infty\rho(\xi)d\xi\int_1^\infty
\frac{\rho(\eta)d\eta}{(\xi^2-\eta^2)}[\xi^2(1+2m_e\xi r)e^{-2m_e\xi
r}- \eta^2(1+2m_e\eta r)e^{-2m_e\eta r}].
\end{displaymath}
Averaging (86) over the wave functions (2), we obtain the following
correction to the interval (82):
\begin{equation}
\Delta E^{fs}_{VP-VP}=\frac{\mu^3\alpha^2(Z\alpha)^4}{288\pi^2m_1^2}\left[1+2a_\mu
+\frac{2m_1}{m_2}(1+a_\mu)\right]\int_1^\infty\rho(\xi)d\xi\times
\end{equation}
\begin{displaymath}
\times\int_1^\infty
\rho(\eta)d\eta\frac{1}{(\xi^2-\eta^2)}\left[\xi^2\frac{6\frac{m_e\xi}{W}+1}{(\frac{2m_e\xi}{W}+1)^3}-
\eta^2\frac{6\frac{m_e\eta}{W}+1}{(\frac{2m_e\eta}{W}+1)^3}\right]=0.000003~meV.
\end{displaymath}
Two-loop vacuum polarization potential and the correction to
fine structure $(2P_{3/2}-2P_{1/2})$ are given by
\begin{equation}
\Delta V_{2-loop~VP}^{fs}(r)=\frac{2Z\alpha^3}{3\pi^2r^3}
[\frac{1+2a_\mu}{4m_1^2}+\frac{1+a_\mu}{2m_1m_2}]\int_0^1
\frac{f(v)dv}{1-v^2}e^{-\frac{2m_er}{\sqrt{1-v^2}}}(1+\frac{2m_er}
{\sqrt{1-v^2}})({\bf L}{\mathstrut\bm\sigma}_1),
\end{equation}
\begin{equation}
\Delta E^{fs}_{2-loop~VP}=\frac{\mu^3\alpha^2(Z\alpha)^4}{48\pi^2m_1^2}
\left[1+2a_\mu+\frac{2m_1}{m_2}(1+a_\mu)\right]\int_0^1\frac{f(v)dv}{1-v^2}
\frac{(6\frac{m_e}{W\sqrt{1-v^2}}+1)}{(1+\frac{2m_e}{W\sqrt{1-v^2}})^3}=0.00002~meV.
\end{equation}

Two-loop vacuum polarization contributions in second order
perturbation theory shown in Fig.4(a,d-f) ($\Delta V^B\to\Delta V^{fs}$),
have the same order $\alpha^2(Z\alpha)^4$. For their calculation it is necessary to
employ the modified Coulomb potential by two-loop vacuum polarization \cite{M2,M3}.
The amplitude in Fig.4(e-f) gives the following correction of order
$\alpha^2(Z\alpha)^4$ to fine structure splitting:
\begin{equation}
\Delta E^{fs}_{2-loop~VP,SOPT}=\frac{\mu^3\alpha^2(Z\alpha)^4}{3\pi^2m_1m_2}\left[1+a_\mu
+\frac{m_2}{2m_1}(1+2a_\mu)\right]\int_0^1\frac{f(v)dv}{1-v^2}\times
\end{equation}
\begin{displaymath}
\times\frac{1}{(1+\frac{2m_e}{W\sqrt{1-v^2}})^6}\left[5\frac{2m_e}{W\sqrt{1-v^2}}+4(1+\frac{2m_e}{W\sqrt{1-v^2}})
\ln(1+\frac{2m_e}{W\sqrt{1-v^2}})\right]
=0.000026~meV.
\end{displaymath}
Two other contributions from amplitudes in Fig.4(a,d) have the similar integral
structure. Their numerical values are included in Table II.

There exists also the correction to fine structure splitting due to nuclear
structure. In $1\gamma$-interaction it is related with the charge
form factor of the deuteron. Fine structure potential
(81) is obtained for the point deuteron. In the case of the deuteron
of finite size we can express the contribution of nuclear
structure to fine structure splitting in terms of the charge radius \cite{ekm2011}:
\begin{equation}
\Delta E^{fs}_{str}=-\frac{\mu^5(Z\alpha)^6}{64m_1^2}r_d^2[1+\frac{2m_1}{m_2}+
2a_\mu(1+\frac{m_1}{m_2})]=-0.00028~meV.
\end{equation}
Earlier the calculation of the nuclear structure corrections to the energies of
P-levels of order $(Z\alpha)^6$ was performed in \cite{Friar1}.
Our numerical result (91) for the fine structure splitting agrees with the
calculation in \cite{Friar1}.

\begin{table}
\caption{Fine structure of $2P$-state in muonic deuterium atom.}
\bigskip
\begin{ruledtabular}
\begin{tabular}{|c|c|c|}   \hline
Contribution to fine & Numerical value & Equation, \\
splitting $\Delta E^{fs}$            &in meV & Reference
\\  \hline
Contribution of order $(Z\alpha)^4$ &              &         \\
$\frac{\mu^3(Z\alpha)^4}{32m_1^2}\left(1+\frac{2m_1}{m_2}\right)$ &
8.83848 &  (82), \cite{BR4,EGS} \\ \hline
Muon AMM contribution &       &           \\
$\frac{\mu^3(Z\alpha)^4}{16m_1^2}a_\mu\left(1+\frac{m_1}{m_2}\right)$
& 0.01957 &(82), \cite{BR4,EGS}  \\  \hline
Contribution of order
$(Z\alpha)^6$ &0.00031    & (82), \cite{BR4,EGS}   \\  \hline
Contribution of order $(Z\alpha)^6m_1/m_2$ &-0.00001 & (82), \cite{BR4,EGS}  \\  \hline
Contribution of order $\alpha(Z\alpha)^4$  &          &        \\
in first order PT $\langle\Delta V^{fs}_{VP}\rangle$ & 0.00346 & (84) \\   \hline
Contribution of order $\alpha(Z\alpha)^4$  &          &        \\
in second order PT   & 0.00229   & (85)   \\
$\langle\Delta V^C_{VP}\cdot\tilde G\cdot \Delta V^{fs}\rangle$ & &  \\   \hline
Contribution of order $\alpha(Z\alpha)^6$  &          &       \\
$\frac{\alpha(Z\alpha)^6\mu^3}{32\pi
m_1^2}\left[\ln\frac{\mu(Z\alpha)^2}{m_1} +\frac{1}{5}\right]$ &-0.00001 &  (82), \cite{EGS} \\   \hline
VP Contribution from $1\gamma$ interaction &      &      \\
of order $\alpha^2(Z\alpha)^4$ $\langle\Delta V^{fs}_{VP-VP}\rangle$
& 0.000003 & (87)   \\   \hline
VP Contribution from $1\gamma$ interaction &      &      \\
of order $\alpha^2(Z\alpha)^4$ $\langle\Delta
V^{fs}_{2-loop,VP}\rangle$    & 0.00002   & (89)   \\   \hline
VP Contribution in second &      &      \\
order PT of order $\alpha^2(Z\alpha)^4$    & 0.000002   & Fig.4(a),~$\Delta V^B\to\Delta V^{fs}$   \\
$\langle\Delta V^C_{VP}\cdot\tilde G\cdot\Delta V^{fs}_{VP}\rangle$    &   &   \\  \hline
VP Contribution in second &      &      \\
order PT of order $\alpha^2(Z\alpha)^4$    & -0.000001   & Fig.4(d),~$\Delta V^B\to\Delta V^{fs}$   \\
$\langle\Delta V^C_{VP-VP}\cdot\tilde G\cdot\Delta V^{fs}\rangle$    &   &   \\  \hline
VP Contribution in second &      &      \\
order PT of order $\alpha^2(Z\alpha)^4$    & 0.000026   & (90),~Fig.4(e-f),~$\Delta V^B\to\Delta V^{fs}$   \\
$\langle\Delta V^C_{2-loop,VP}\cdot\tilde G\cdot\Delta
V^{fs}\rangle$    & &   \\  \hline
Nuclear structure correction & -0.00028  &   (91), \cite{Friar1} \\
in $1\gamma$ interaction &   &   \\   \hline
Summary contribution  &8.86386&    \\   \hline
\end{tabular}
\end{ruledtabular}
\end{table}

\section{Summary and conclusion}

In this work, various corrections of orders $\alpha^3$, $\alpha^4$,
$\alpha^5$ and $\alpha^6$ are calculated to the Lamb shift
$(2P_{1/2}-2S_{1/2})$ and fine structure splitting $(2P_{3/2}-2P_{1/2})$ in muonic deuterium atom.
Contrary to earlier
performed investigations of the energy spectra of light muonic atoms
in \cite{BR3,BR4,BR1}, we have used the three-dimensional
quasipotential approach for the description of two-particle bound
state. Our analysis of different contributions to the Lamb shift
accounts for the terms of two groups. First group contains the
specific corrections for muonic deuterium, connected with the
electron vacuum polarization effects, nuclear structure and recoil
effects in first and second order perturbation theory. As a rule the
contributions of this group are obtained in integral
form over auxiliary parameters and calculated numerically.
The necessary order corrections of second group include
analytical results known from the corresponding calculation in
the electronic hydrogen Lamb shift. Recent advances in the physics
of the energy spectra of simple atoms are presented in the review
articles \cite{EGS,SY,SGK} which we use in this study. Numerical values
of all corrections are written in Tables I, II, which contain also
basic references on the earlier performed investigations (other
references can be found in Ref.\cite{BR3,BR4,EGS}). We compare our intermediate
results for different corrections with calculations given in \cite{BR4}.
Most part of the results including the Uehling, K\"allen-Sabry, Wichmann-Kroll
corrections, muon Lamb shift contribution, nuclear size and VP corrections
and recoil terms agree well.
Our results for relativistic contributions to
the vacuum polarization are in agreement with those obtained in \cite{jentschura3}.
Second order VP correction (39) and (40) agrees with the result of \cite{KIKS} just as
the three loop VP contribution which is determined in Table I by three lines
corresponding to one-photon interaction (0.0060 meV), second order PT (0.0025 meV)
and third order PT (0.0001 meV).
Total numerical value $202.4139$ meV of the Lamb shift $(2P-2S)$ in muonic deuterium
atom from Table I is in good agreement with the theoretical result
$202.263$ meV obtained in \cite{BR4}. The difference of our
result from Ref.\cite{BR4} is connected with the
calculation of new contributions of higher order in $\alpha$ and $m_1/m_2$, the
proton structure and polarizability correction \cite{Pachucki2011}
and slightly different numerical value of the charge radius of the deuteron
$r_d$ used in this work. Two-loop vacuum polarization contribution 0.1720 meV of
order $\alpha^2(Z\alpha)^2$ in second order PT is absent in \cite{BR4}.
The value of the charge radius $r_d=2.139(3)$ fm is used in \cite{BR4}.
Fine structure splitting $(2P_{3/2}-2P_{1/2})$ 8.86386 meV in Table II agrees also with
the result 8.864 meV from \cite{BR4}. Recently, improved analysis of different corrections
to the Lamb shift in $(\mu d)$ is performed in \cite{borie2011}. Total value of the
Lamb shift $(2P_{1/2}-2S_{1/2})$ for $r_d=2.130$ fm according to Table 4 from \cite{borie2011}
amounts 202.9440 meV. This value exceeds our result 202.7375 meV by 0.2065 meV. In our opinion
the only two essential differences between our Table I and \cite{borie2011} are related
with the Zemach correction 0.4329 meV and polarizability correction 1.5 meV \cite{borie2011}.
It was shown in \cite{Pachucki2011} that the Zemach correction is canceled by the
deuteron excited states contribution. As a result the nuclear structure and polarizability
contribution is equal to 1.680 meV \cite{Pachucki2011} which we use in our work.

As has been mentioned above numerical values of corrections are obtained
with an accuracy $0.0001$ meV because certain contributions to the Lamb shift $(2P-2S)$
of order $\alpha^6$ attain the value of tenth part of $\mu eV$. The
theoretical error caused by uncertainties in fundamental
parameters (fine structure constant, particle masses) entering the
leading order contributions is around $10^{-5}$ meV. The other part of
theoretical error is related to the QED corrections of higher order. This part can
be estimated from the leading contribution of higher order in
$\alpha$: $m_1\alpha(Z\alpha)^6\ln(Z\alpha)/\pi n^3\approx 0.0001$
meV. Theoretical uncertainty connected with nuclear structure and polarizability
contributions is equal to $0.0160$ meV
\cite{Pachucki2011}. We have also small theoretical uncertainty determined by
HVP contribution which we estimate in $5\%$ ($\pm 0.0006~meV$). This estimation is
based on the experimental uncertainty in the cross section of $e^+e^-$ annihilation
into hadrons.
The rounding errors can amount to $0.0001\div 0.0002~meV$.
Finally, the biggest theoretical error $\pm 0.0550$ meV (for $r_d=0.1424(21)$ fm)
is related to the uncertainty of the deuteron charge radius. Thereby, the
total theoretical error of the calculation is equal to $\pm 0.0573$ meV. To obtain
this estimate we add the above mentioned uncertainties in quadrature.

Let us summarize the basic particularities of the Lamb shift calculation
performed above.

1. Numerical value of specific parameter $m_e/\mu Z\alpha=0.7$
in muonic deuterium atom is sufficiently large, so the
electron vacuum polarization effects play essential role in the
interaction of the bound particles. We have considered the one-loop, two-loop and
three-loop VP contributions to the Lamb shift $(2P_{1/2}-2S_{1/2})$. A number of
important vacuum polarization contributions from $1\gamma$-interaction agrees
with the results obtained in \cite{BR4,KN1,KN2,KIKS}.

2. Nuclear structure effects are expressed in the Lamb shift of
muonic deuterium atom in terms of the deuteron charge radius $r_d$.
We analyze complex effects due to nuclear structure and vacuum polarization
in first and second orders of perturbation theory.
The elastic nuclear structure contribution from two-photon exchange amplitudes
is canceled by the deuteron polarizability correction \cite{Pachucki2011}.

3. Nuclear structure and polarizability effects give the largest theoretical
uncertainty in the total value of the Lamb shift $(2P-2S)$. It is useful to express
the final theoretical value of the $(2P-2S)$
Lamb shift in the form $\Delta E^{Ls}(2P-2S)=(230.4511-6.108485\cdot r_d^2$) meV
with the value of the deuteron charge radius defined in fm. Then, comparing this
expression with the experimental value of the Lamb shift measured with the precision
0.01 meV (50 ppm) we can obtain more accurate value of $r_d$ with an accuracy $0.0005$ fm.

\acknowledgments
The authors are grateful to A. Antognini, R.N. Faustov, F. Kottmann,
T. Nebel, R. Pohl for the information about experimental results with muonic
hydrogen, new muonic helium project of CREMA collaboration  and useful
discussions. We thank U.D. Jentschura for the reading our manuscript,
valuable remarks about the recoil correction (64) and comments to our calculation.
This work is supported by the Russian Foundation for Basic Research
(grant No. 11-02-00019) and the Federal Program "Scientific and pedagogical
personnel of innovative Russia"(grant No. NK-20P/1).

\end{document}